# Healthy diets are affordable but often displaced by other foods in Indonesia

Leah Costlow[a,b*], Rachel Gilbert[c], William A. Masters[c,d], Flaminia Ortenzi[e], Ty Beal[f], Ashish Deo[g], Widya Sutiyo[h], Sutamara Noor[h], Wendy Gonzalez[e]

* Contact: costlow1@msu.edu

**Author affiliations:**
[a] Department of Agricultural, Food, and Resource Economics, Michigan State University, East Lansing MI, USA
[b] Department of Food Science and Human Nutrition, Michigan State University, East Lansing MI, USA
[c] Friedman School of Nutrition Science and Policy, Tufts University, Boston MA, USA
[d] Department of Economics, Tufts University, Medford MA, USA
[e] Global Alliance for Improved Nutrition, Geneva, Switzerland
[f] Global Alliance for Improved Nutrition, Washington DC, USA
[g] Global Alliance for Improved Nutrition, London, UK
[h] Global Alliance for Improved Nutrition, Jakarta, Indonesia

**Acknowledgements:**
This work was funded by the Dutch Ministry of Foreign Affairs, using methods developed through the Food Prices for Nutrition project at Tufts University supported by the Bill & Melinda Gates Foundation. This work was done while the first author was a doctoral student in the Friedman School of Nutrition Science and Policy at Tufts University.

**Author contributions: CRediT**
**Leah Costlow:** Data curation, Formal analysis, Writing – review & editing, Writing – original draft, Conceptualization. **Rachel Gilbert:** Data curation, Formal analysis, Writing – review & editing, Conceptualization. **William Masters:** Writing – review & editing, Conceptualization, Methodology. **Flaminia Ortenzi:** Writing – review & editing, Conceptualization, Project administration. **Ty Beal:** Writing – review & editing, Conceptualization, Methodology. **Ashish Deo:** Writing – review & editing. **Widya Sutiyo:** Writing – review & editing, Resources. **Sutamara Noor:** Writing – review & editing, Resources. **Wendy Gonzalez:** Writing – review & editing, Conceptualization, Project administration.

**Keywords:**
Consumer prices; Diet costs; Diet affordability; Diet quality; Food environments; Nutrition

## Highlights

- In places where healthy diets are affordable, other factors are responsible for poor diets.
- We use Indonesia as a case study to demonstrate a 3-phase approach for analyzing barriers to consumption of healthy diets.
- In Indonesia, healthy diets are affordable but not chosen while consumption of unhealthy discretionary foods is high.
- Nutrient and food group adequacy are low for households across all income groups.
- Policymakers can use this approach to understand whether the cost and affordability of healthy diets are primary drivers of low dietary adequacy.

## Abstract

New methods for modeling least-cost diets that meet nutritional requirements for health have emerged as important tools for informing nutrition policy and programming around the world. This study develops a three-step approach using cost of healthy diet to inform targeted nutrition programming in Indonesia. We combine detailed retail prices and household survey data from Indonesia to describe how reported consumption and expenditure patterns across all levels of household income diverge from least cost healthy diets using items from nearby markets. In this analysis, we examine regional price variations, identify households with insufficient income for healthy diets, and analyze the nutrient adequacy of reported consumption patterns. We find that household food spending was sufficient to meet national dietary guidelines using the least expensive locally available items for over 98% of Indonesians, but almost all households consume substantial quantities of discretionary foods and mixed dishes while consuming too little energy from fruits, vegetables, and legumes, nuts, and seeds. Households with higher incomes have higher nutrient adequacy and are closer to meeting local dietary guidelines, but still fall short of recommendations. These findings shed new light on how actual food demand differs from least-cost healthy diets, due to factors other than affordability, such as taste, convenience, and aspirations shaped by marketing and other sociocultural influences.

## 1. Background and motivation

Recent methods for modeling least-cost diets that meet nutritional requirements for health have expanded on traditional food security metrics and provided new insights into food security around the world (Masters et al., 2025). These diet modeling methods, known as the Cost and Affordability of a Healthy Diet (CoAHD) indicators, were developed in a series of country studies in 2018-19 and have been used for global monitoring by FAO and the World Bank since 2020. (FAO, IFAD, UNICEF, WFP and WHO, 2020, 2021, 2022, 2023, 2024). These metrics provide a benchmark with which to distinguish a population's access to healthy diets from their actual food choices, thereby guiding action to reduce malnutrition and diet-related disease (Herforth et al., 2025). A variety of global analyses have used these methods to summarize the extent to which low-income people around the world cannot afford healthy diets, and national governments have adopted the CoAHD indicators for country-level poverty monitoring (Gilbert et al., 2024; Wallingford et al., 2024; Herforth et al., 2025), with a focus on improving the affordability of healthy diets through supply interventions to reduce costs or consumer interventions to support household incomes. Poverty reduction is critically important for improving food security and wellbeing for vulnerable people around the world. However, in many countries diet quality is poor even when least-cost healthy diets are affordable due to excess consumption of unhealthy foods and/or inadequate intake of healthy food groups (Masters et al., 2025). While global monitoring of CoAHD reveals that about 3 billion people cannot afford a healthy diet, the remaining 5 billion could afford least-cost options but often consume other foods instead (FAO, IFAD, UNICEF, WFP and WHO, 2024). Least-cost diet modeling methods have not yet been used to their full potential for describing food choice relative to nutritional benchmarks in places where incomes are not a primary constraint on the consumption of healthy diets.

The objective of this study is to develop and conduct a three-step approach using CoAHD to inform the design of targeted nutrition programming. We demonstrate this new approach in an upper-middle income country where absolute monetary cost is not the primary constraint on healthy diets. However, the methods shown here are adaptable to countries across the income spectrum. We begin by measuring the degree to which parts of the country face high prices for even the lowest cost options, such that innovations or policy changes could bring prices down to the frontier of what is available in other regions with comparable agroecological, economic, and social conditions. We then identify which households cannot afford a healthy diet, for which income growth or expanded social protection would be needed to bring healthy diets within reach. Finally, we describe the nutrient and food group adequacy of reported diets, considering the extent to which low-cost options for a healthy diet are displaced by other items, resulting in insufficient consumption of foods needed for health such as vegetables and fruits, as well as excess consumption of foods associated with diet-related diseases such as items high in added sugar, salt, and saturated fat.

We apply this approach to Indonesia as an ideal case study given its rapid economic development and ongoing nutrition transition. Indonesia is an upper-middle-income country with wide variation in household income and living standards among its 38 provinces and across the urban-rural continuum. The share of people living in extreme poverty at below $2.15 per day has declined dramatically from 74.3% in 1984 to 1.8% in 2023 (World Bank, 2024). The share of people living below the national poverty line of about $3.68 per day has declined more slowly, falling from 17.8% in 2006 before hovering around 10% since 2018 (BPS, 2023a). In many countries around the world, growth and development has improved quality of life in many ways but has had mixed consequences for diet quality, as people choose to consume more of some health-promoting foods but also increase consumption of health-harming foods that may be aspirational, convenient, or preferred for other reasons (Miller et al., 2022). Recent studies have described Indonesia's ongoing dietary transition, with shifts among food groups towards more animal source foods, refined grains, edible oil and sweeteners, as well as a shift in food sources towards packaged foods and meals away from home (Mehraban & Ickowitz, 2021; Nurhasan et al., 2024). This shift may be linked to increased rates of overweight, obesity, hypertension, diabetes, and other diet-related diseases through mechanisms including increased consumption of ultra-processed foods (Pagliai et al., 2021) and through increased consumption of energy, sodium, added sugars, unhealthy fats, and other dietary risk factors independent of food processing level.

Previous studies using least-cost diet calculations have either focused on low-income countries or have used large global datasets for cross-country comparisons (Gilbert et al., 2024; Herforth et al., 2025; Masters et al., 2025; Wallingford et al., 2024). One study from the World Food Programme investigated the cost of diets that meet nutrient requirements but did not systematically incorporate food-based recommendations for health (WFP & Kementrerian PPN/Bappenas, 2017). Other work has compared national food availability in Indonesia with a range of nutritional benchmarks (de Pee et al., 2021). No study has yet examined least-cost healthy diets in a middle-income country where affordability may not pose a significant barrier to consumption of healthy diets. This paper complements the existing literature on least-cost healthy diets and on the nutrition transition in Indonesia by demonstrating how the CoAHD methodology can be used in middle-income countries. Comparing the dietary adequacy of reported consumption with least-cost healthy diets provides further insights into the nature of the nutrition interventions that are necessary in this context.

## 2. Data and methods

Each of the three phases of this study incorporates different data sources as benchmarks for describing and interpreting reported household data. In the first phase, food group costs and the total cost of a healthy diet are compared across geographic locations to understand geographic variation in diet costs. In the second phase, the cost of a healthy diet is compared with household food spending to assess affordability by examining whether household income is sufficient to purchase healthy diets . In the third and final phase, we draw on different nutritional benchmarks to model the alignment of reported diets with needs for nutrient adequacy and health. Throughout the analysis, we use Indonesia's extensive cross-sectional household survey data to investigate how dietary patterns vary with income. Income is measured using household expenditure, following the usual practice recommended by Deaton (1997) for situations where earnings are highly variable and difficult to measure over time. In these settings, expenditures provide the most accurate measure of the household's long run average earnings and hence what they can usually afford to buy, which can be especially useful for analysis of food and nutrition data in low- and middle-income countries (Fiedler et al. 2012).

We calculate modeled diet costs using food price data collected by Badan Pusat Statistik (BPS) for the calculation of Indonesia's Consumer Price Index (CPI) (BPS, 2023b). BPS tracks the availability and price of 193 commonly sold food and beverage items each month, 79 of which are condiments and spices, beverages, infant foods, or snack items, which are excluded from our analysis. The remaining 114 items are distributed across the seven food groups specified in the country's national dietary guidelines or are classified as mixed dishes that cannot be definitively assigned to a single recommended food group. Reported diets for March 2023 are derived from Indonesia's national household survey, the National Socio-Economic Survey (Susenas), which collected data from 341,802 households and provides representative samples of the population at the national, provincial, and district or municipality levels (BPS, 2023c). We use household sampling weights from Susenas to adjust all estimates for reported consumption, reported expenditures, and modeled diet costs.

We calculate the cost of modeled diets using the CoHD method as described in Herforth et al., 2022 and Herforth et al., 2025. Using this simple rank-order optimization method, we identify the least-cost items within each recommended food group at each of 514 unique locations, and calculate the total cost to purchase the recommended daily quantity from all food groups to meet the needs of a representative person, defined as a healthy adult woman. We convert all item prices from cost per weight to cost per kilocalorie using our original matching of item descriptions to food composition data. This conversion allows for accurate substitutions between items within each food group without distortions from water weight.

After computing least-cost healthy diets, we compare the cost and composition of these modeled diets to the cost and composition of diets reported through Susenas. The survey's consumption module asks households to report expenditures and quantities consumed for 178 food and beverage products, resulting in a total of over 17 million household-food observations. We follow standard practice in nutritional epidemiology by using energy-adjustment, in which we preserve the ratios of reported item-level consumption for each household while converting the total energy to a reference level of 2330 kilocalories per female adult equivalent. This approach reflects the consensus that survey responses systematically understate the total quantity of all foods consumed, but have no known bias in terms of energy shares (Willett, 2012). Energy adjustment allows quantities to be scaled up or down based on other evidence about energy balance for each individual, with adherence to dietary guidelines measured in aggregate on average over all members of the household.

In the next phase of the analysis, we estimate the affordability of healthy diets by comparing each household's energy-adjusted food spending per adult equivalent with the cost of a healthy diet. Households are categorized as unable to afford a healthy diet if the cost of a healthy diet in their place of residence exceeds their reported food spending. Average rates of unaffordability at the provincial or city level are calculated using household survey sampling weights.

In the final phase of the analysis, we use different benchmarks to describe dietary adequacy with respect to nutrients and food-based dietary guidelines. To describe food group adequacy, we use Indonesia's 2014 national dietary guidelines (Table 1), which also serve as the foundation for our modeled least-cost diet calculations. These guidelines specify the number of items and quantities needed from each of six food groups needed for long-term health, plus an allowance for a seventh discretionary food group of items that would not be harmful if consumed in sufficiently small quantities. The discretionary food group includes sweets, salty snacks, instant noodles, and processed meats. Food requirements depend on household composition, with diet costs expressed in adult equivalent terms based on the age and sex of each individual listed in the household roster. To align with global monitoring standards, our adult-equivalent reference person is based on WHO data for a 30-year-old woman at 2,330 calories per day (Schneider & Herforth, 2020).

Adherence to dietary guideline recommendations in terms of food categories is shown using the mean food group adequacy (MFGA) for each household, following the method introduced by Costlow et al. (2025). For each recommended food group, we score consumption per adult equivalent as a fraction of the reference level, capped at one to express the degree to which each target is met. Mean adequacy is the average across the six food groups needed for a healthy diet, omitting the discretionary foods allowance. By construction, MFGA for the benchmark modeled diet is 100%, while actual diets fall below this benchmark when they fail to meet targets for some food groups needed for health, regardless of whether or not they exceed targets for other food groups. Unhealthy diets may be unbalanced in different ways, each contributing to particular forms of malnutrition and diet-related disease. In contrast, diets recommended in dietary guidelines tend to have similar balance among food groups, in that food items that form part of a healthy diet can substitute freely within groups while complementing the other food groups. Since each food group is equally important for a balanced diet, an equally weighted level of average adequacy across the six required groups provides a simple representation of how foods are used inside the body to sustain the population's lifelong health.

**Table 1. Indonesian dietary guidelines and number of items in available price data**

| **Food groups** | **Number of items** | **Quantity (kcal/day)** | **Calorie share (%)** | ***Number of items in food price list*** |
|---|---|---|---|---|
| Starchy staples | 2 | 1256 | 54 | *11* |
| Oils and fats | 1 | 275 | 12 | *1* |
| Fruits | 2 | 138 | 6 | *16* |
| Vegetables | 3 | 97 | 4 | *16* |
| Legumes nuts and seeds | 1 | 265 | 11 | *6* |
| Animal source foods | 2 | 199 | 9 | *54* |
| Discretionary foods | 1 | 100 | 4 | *10* |
| Excluded | 0 | 0 | 0 | *79* |
| Total | 12 | 2,330 | 100 | *193* |

*Source: Quantity targets specified in the Indonesian dietary guidelines are from Herforth et al., 2025; analysis of items used by BPS for the Indonesian consumer price index is the authors' calculations for this study.*

Adherence to nutrient requirements is shown using mean nutrient adequacy (MNA) as established in Madder & Yoder (1972). We identify the composition of each food in the analysis by matching item descriptions to food composition databases, relying on USDA's flagship SR28 database while supplementing with regional databases for a small number of items as needed (USDA Agricultural Research Service, 2016; Kementerian Kesehatan Republik Indonesia, 2017; Hulshof et al., 2019). Food composition tables support the calculation of the cost of a healthy diet by providing information on energy density and edible portions, which enables the conversion of prices into kilocalorie terms. This also allows us to describe the nutrient adequacy of both modeled least-cost healthy diets and reported diets by comparing total dietary nutrient content with recommended nutrient intakes (RNIs). For micronutrients, reference intakes are taken from either the Recommended Daily Allowance (RDA) from IOM or from the Population Reference Intake (PRI) from EFSA (Allen et al., 2020). Micronutrients included in the analysis are calcium, iron, zinc, thiamin, riboflavin, niacin, vitamin B6, vitamin B12, vitamin C, folate, and vitamin A. For protein, lipids, and carbohydrates, recommended minimum intakes are taken from the lower bound of the Acceptable Macronutrient Distribution Range (AMDR) (Trumbo et al., 2002). We calculate the nutrient adequacy ratio (NAR) for each nutrient as the quantity of the nutrient amount consumed or available in the modeled diet for consumption divided by the RNI for that nutrient, where NAR is capped at 1. The mean nutrient adequacy (MNA) of the diet is calculated as the average NAR across all nutrients in the reported or modeled diet, resulting in a value ranging from 0 to 1 where 1 indicates that all nutrient requirements are met.

## 3. Results

### *3.1. Phases 1 and 2: Cost and affordability of a healthy diet*

Matching each individual household's income to diet costs at their location reveals that about 1.5% of all Indonesian households cannot afford a healthy diet (Table 2). Within the poorest income quintile, just 5% of households cannot afford a healthy diet, while nearly all households in the remaining four income quintiles can afford a healthy diet. Households in the lowest income quintile have larger households, are more likely to reside in rural areas, and devote substantially more of their incomes on food as compared to the wealthiest households.

**Table 2. Descriptive statistics by household income quintile**

| | Quintiles at National Level | | | | | | |
|---|---|---|---|---|---|---|---|
| | 1st Quintile | 2nd Quintile | 3rd Quintile | 4th Quintile | 5th Quintile | Total | Test |
| N | 65,152,371 (25.6%) | 56,190,682 (22.1%) | 49,689,435 (19.5%) | 44,959,730 (17.7%) | 38,621,039 (15.2%) | 254,613,257 (100.0%) | |
| # of household members | 4.98 (1.68) | 4.55 (1.53) | 4.27 (1.45) | 4.05 (1.41) | 3.77 (1.37) | 4.40 (1.57) | <0.001 |
| Rural (0/1) | 0.55 (0.50) | 0.48 (0.50) | 0.44 (0.50) | 0.35 (0.48) | 0.23 (0.42) | 0.43 (0.49) | <0.001 |
| Household share of expenditure on food (%) | 62.12 (9.15) | 59.11 (9.68) | 56.59 (10.50) | 52.77 (11.57) | 42.22 (14.22) | 55.71 (12.66) | <0.001 |
| Share of households that cannot afford a healthy diet (%) | 5.02 (21.83) | 0.65 (8.03) | 0.27 (5.16) | 0.07 (2.59) | 0.01 (0.86) | 1.49 (12.12) | <0.001 |

*Note: Standard deviations shown in parentheses. Household sampling weights used for estimation.*

The minimum expenditure needed to purchase a healthy diet as defined by Indonesia's food-based dietary guidelines ranged from 5,600 to 39,000 IDR per adult equivalent per day in March 2023, with a national average cost of 10,503 IDR. Costs are highest by a large margin in inland Papua and on the border with Papua New Guinea, and are also relatively high in parts of Kalimantan and West Java (Figure 1).

**Figure 1. Cost and affordability of a healthy diet by regency and city, March 2023**

**Panel A. Cost of a healthy diet (IDR per day)**

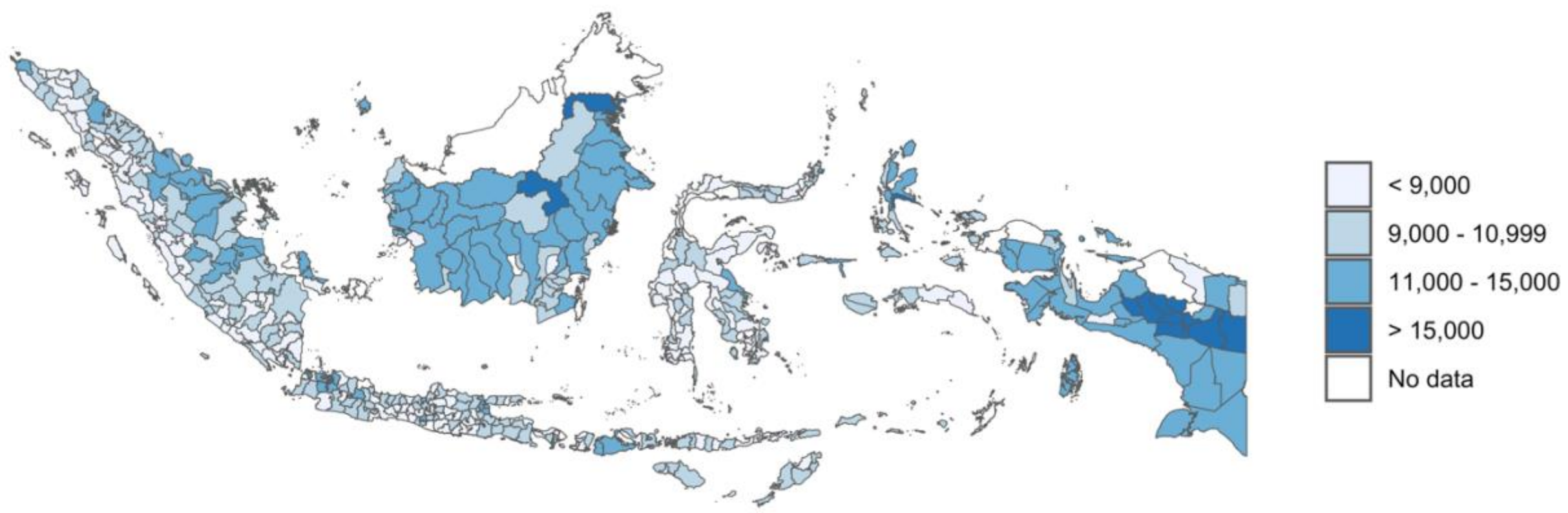

**Panel B. Percentage of households that cannot afford a healthy diet**

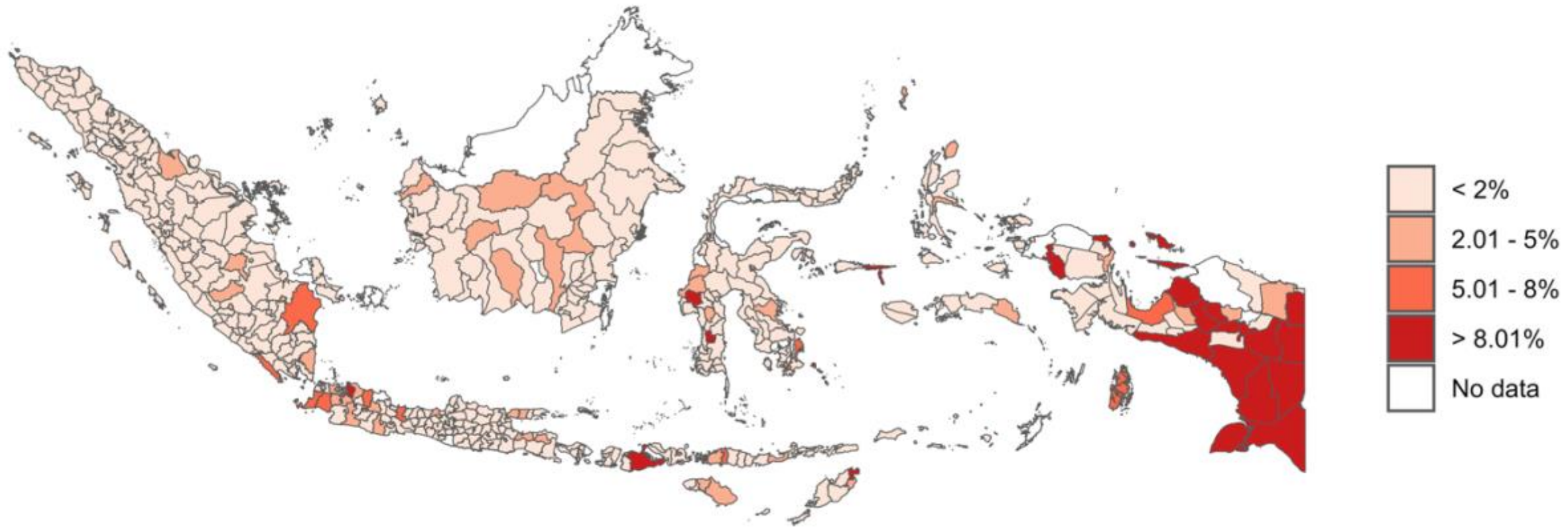

*Note: Data shown in Panel A are the average cost of a healthy diet by regency and city, shown in IDR per person per day. Diet costs are calculated per adult equivalent to provide 2330 kilocalories, meeting the needs of an active 30-year-old woman. Data shown in Panel B are the percentage of households in each regency and city whose reported food spending per person is less than the local cost of a healthy diet. Household sampling weights used for estimation.*

Comparing the two panels of Figure 1 shows how the geographic pattern of high diet costs differs from the pattern of unaffordability, due to low incomes in some places with low food prices and vice versa. This absolute benchmark definition of affordability reveals how access to healthy diets differs around the country, comparing least-cost options to each household's reported spending on food. Areas of high affordability include population centers such as Jakarta, large parts of Kalimantan, and large areas of Sumatera, but affordability varies even within these provinces. Areas of low affordability include places with low population density, notably Papua which has the largest share of households that cannot afford a healthy diet at 20 percent, with unaffordability prevalence as high as 45% in some areas. One reason for this is the unusually high cost of even the least expensive options in Papua, primarily due to high prices for starchy staples, fruits, and vegetables as shown at the far right of Figure 2.

**Figure 2. Average cost (IDR) per day for food groups recommended for a healthy diet and discretionary foods, March 2023**

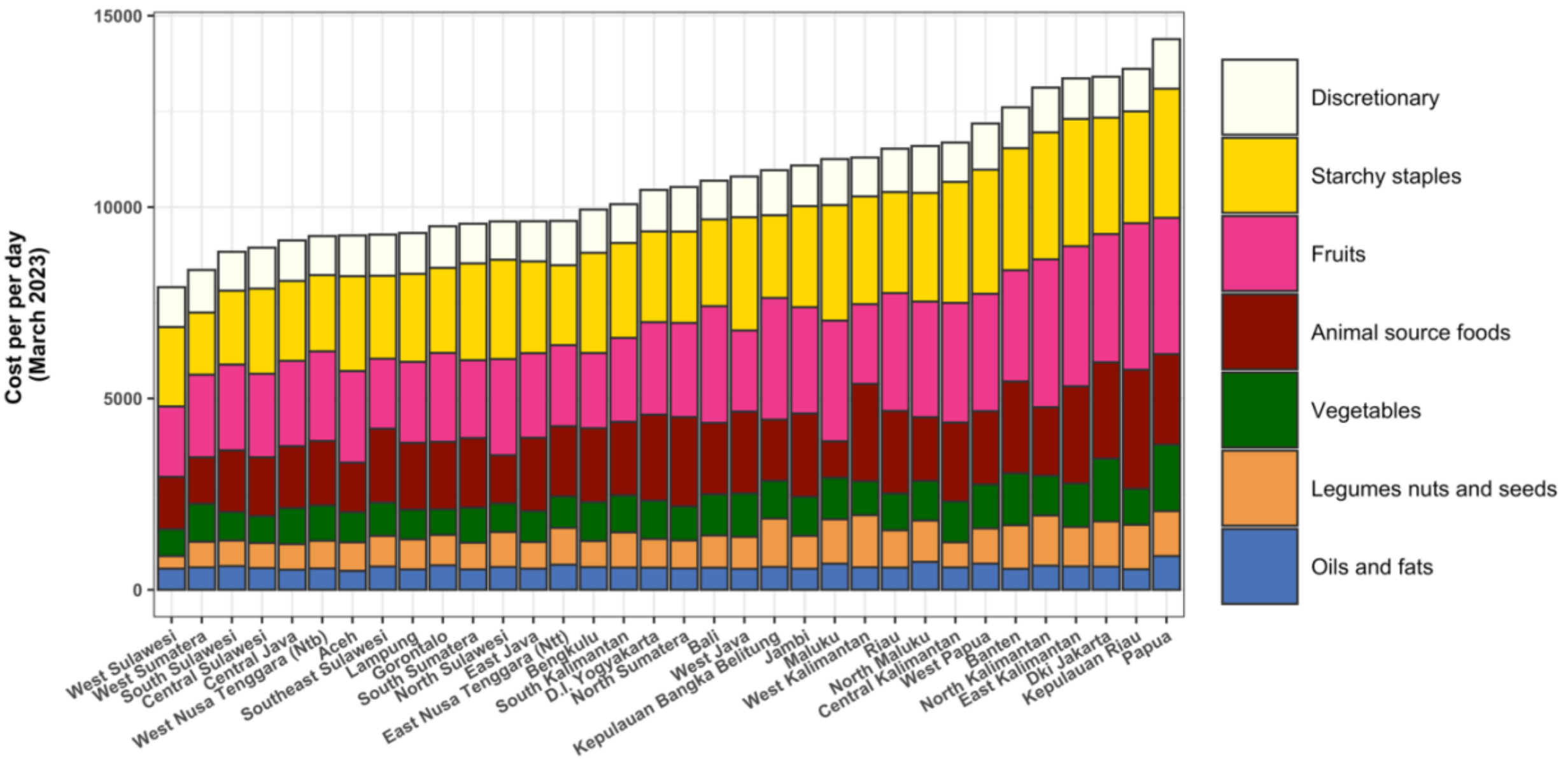

*Note: Data shown are provincial averages for least-cost items meeting Indonesia's dietary guidelines by food group. Weighted average costs are calculated using household sampling weights.*

Costs to meet recommended levels for each food group vary by province, with cost differences driven primarily by starchy staples and fruits that each account for between 20 and 30 percent of the total diet cost, as well as by wide variation in costs for animal source foods. Prevailing prices for starchy staples have a large influence on diet costs due to the large quantities needed (54% of dietary energy), while prices for fruit are influential despite the low energy requirement because of high prices per kilocalorie. A notable feature of Indonesia's food economy is the lower cost share of vegetables and animal source foods in the benchmark modeled diets shown in Figure 2. However, wide geographic variation in the cost of animal source foods nonetheless contributes to overall variation in total diet cost.

Reported food spending generally exceeds each household's least cost benchmark modeled diet, even for those in the lowest income quintiles, as shown in Figure 3. Households in higher income quintiles spend much more on food in absolute terms, primarily driven by larger quantities of animal source foods and mixed dishes. However, all households, even those identified as unable to afford a healthy diet, report meaningful food spending on discretionary foods, which are permitted in moderation but not required as part of a healthy diet, and on mixed dishes. Which are composed of ingredients from multiple food groups. Across all 5 income quintiles, households typically spend more than would be needed to meet their requirements for starchy staples, vegetables, and animal source foods. For example, households in the first (poorest) quintile spend an average of 2,570 IDR per person per day on animal source foods, when the least-cost animal source food items would cost an average of only 1,799 IDR. Spending on starchy staples, legumes, nuts and seeds, and oils and fats is relatively constant across the income spectrum, while spending on other food groups grows substantially as income rises. However, spending across all income groups is less than the healthy diet benchmark for fruits and for legumes, nuts, and seeds.

**Figure 3. Average energy-adjusted expenditures by food group and income quintile vs. cost of a least-cost healthy diet**

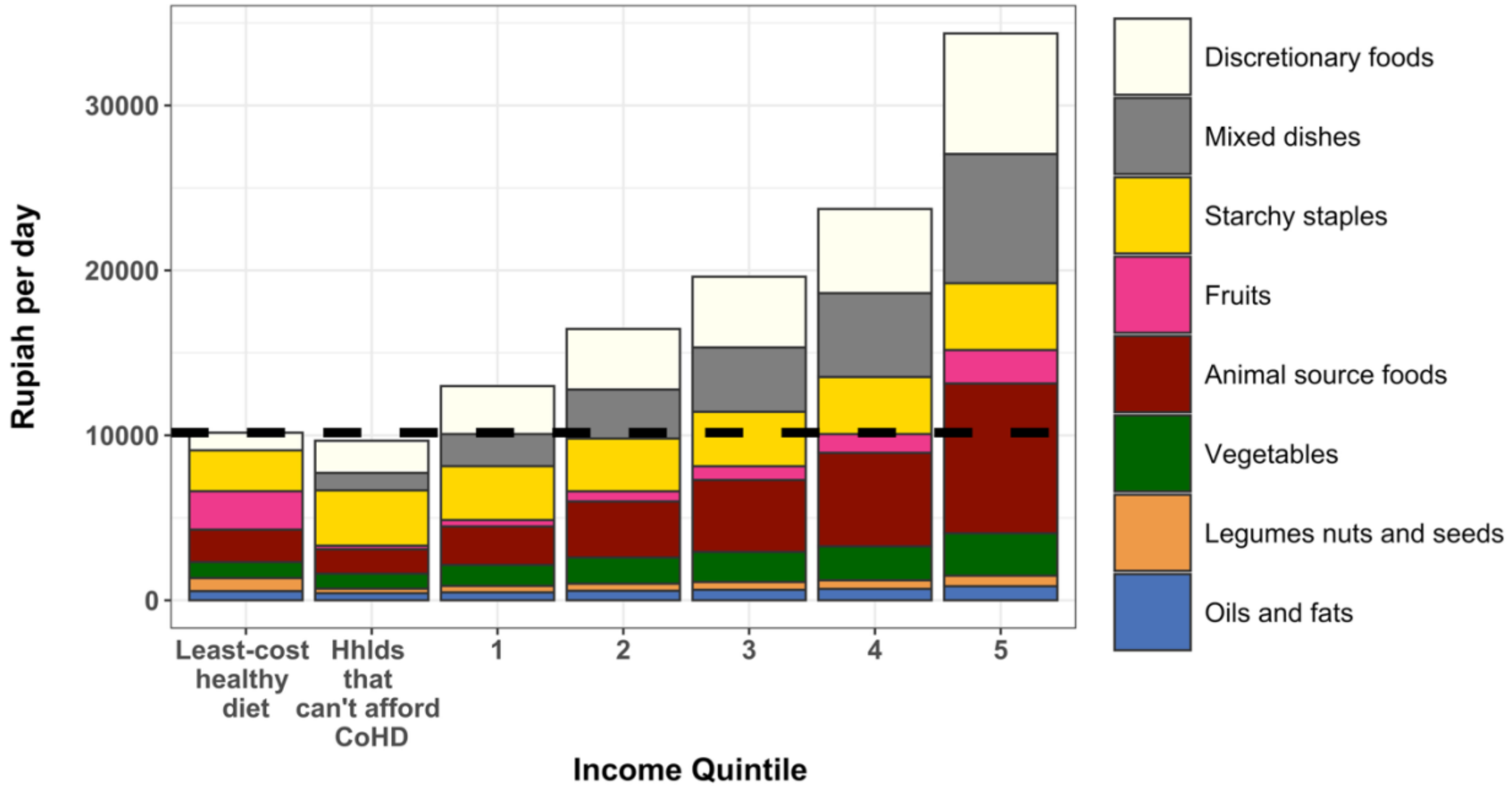

*Note: Data shown are average cost per day for a least-cost healthy diet (left-most column) and average reported household spending by food group and income group (remaining columns). Reported spending is energy-adjusted to 2330 kcals. Weighted average costs and spending levels are calculated using household sampling weights. Black dashed line indicates average CoHD. Households that cannot afford CoHD are those whose reported food spending was less than CoHD after energy adjustment. Quintile 1 includes these households.*

### *3.2. Phase 3 (a): Nutrient and dietary adequacy of reported consumption*

Reported food consumption in Indonesia diverges from national guidelines for most food groups, as shown in Figure 4. Consumption of discretionary foods is extremely high at more than three times the recommended level, while consumption of animal source foods and oils & fats is close to the recommended level. In contrast, there are substantial shortfalls in consumption of vegetables, fruits, and legumes, nuts, and seeds across all income quintiles. Discretionary foods account for 24% of consumed energy, far greater than the recommended energy share of about 4%. In other words, about 20% of dietary energy is diverted from foods that are recommended as part of a healthy diet toward foods that are tolerated only in small quantities. Households that cannot afford CoHD diverge from these patterns in several important ways. They consume a larger amount of energy from starchy staples, exceeding the reference levels by about 150 kcals, and too little energy from animal source foods, with a shortfall of about 50 kcals. However, these households still consume a large share of energy from discretionary foods and mixed dishes, while reported consumption of fruits, vegetables, and legumes, nuts, and seeds remains very low.

**Figure 4. Average energy-adjusted food consumption in Indonesia by income quintile vs. national dietary guidelines**

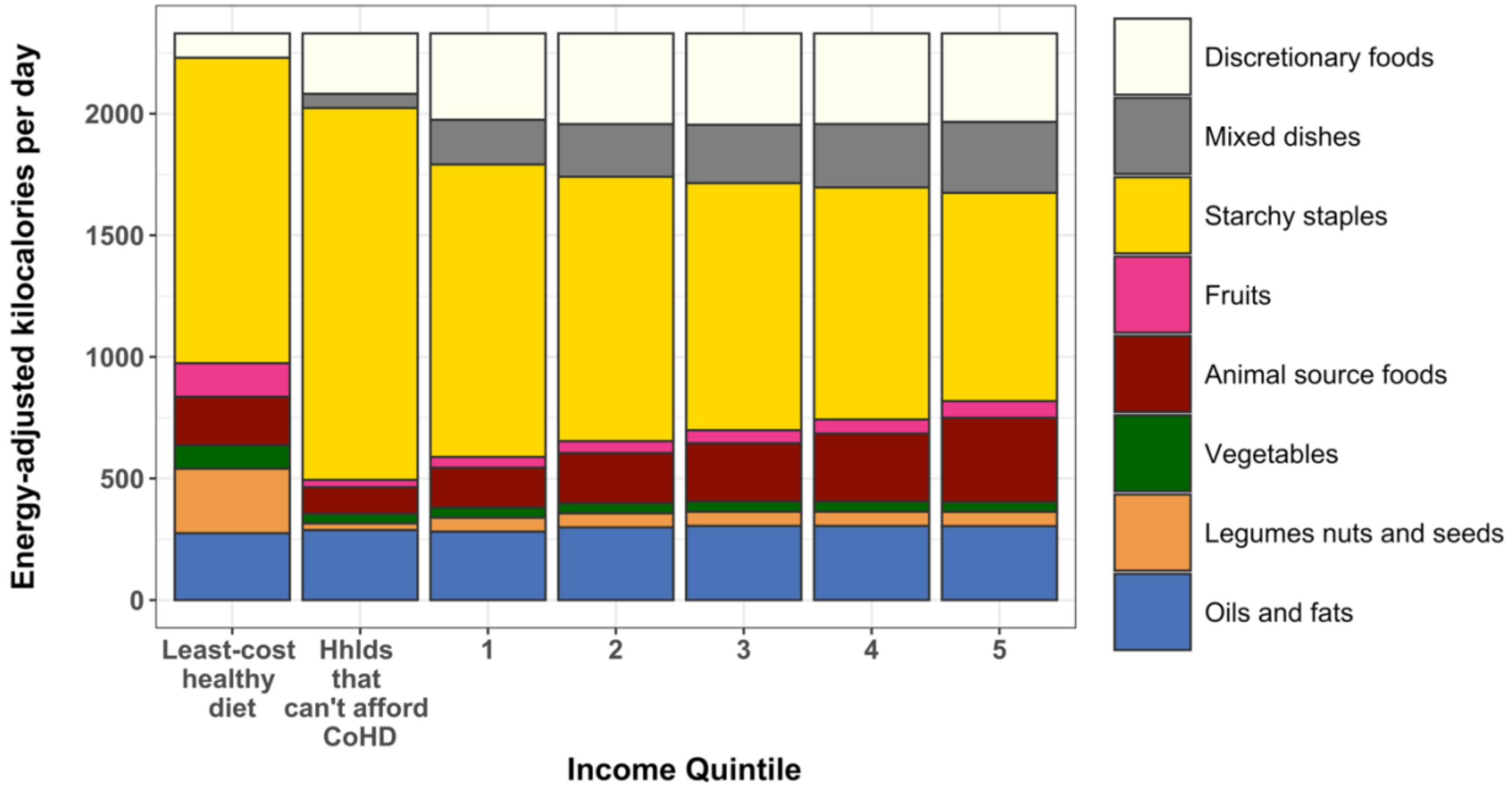

*Note: Data shown are average reported household energy consumption by food group. Reported consumption is energy-adjusted to 2330 kcals to account for measurement error. Averages are calculated using household sampling weights. Households that cannot afford CoHD are those whose reported food spending was less than CoHD after energy adjustment. Quintile 1 includes these households.*

Across all income groups, including households that cannot afford to buy a healthy diet, a large share of consumption comes from mixed dishes. This category includes a diverse range of mixed dishes, which may include recommended food groups that are otherwise inadequate in the diet, including fruits, vegetables, and legumes, nuts, and seeds. However, other mixed dishes contain processed meats, fried foods, or other ingredients that would not be included in a recommended diet. Consumption of animal source foods, discretionary foods and mixed dishes appears to rise with income, while consumption of starchy staples declines, potentially reflecting the conversion of some starchy staples consumption from less processed grains to ready-to-eat meals.

### *3.3. Phase 3 (b): Nutrient and dietary adequacy of least-cost healthy diets vs. reported consumption*

We compare reported consumption of nutrients (Figure 5A) and recommended food groups (Figure 5B) to each respective reference intake to estimate shortfalls and excesses prior to calculating mean adequacy ratios. The observed ranges of nutrient adequacy vary widely with household expenditures, especially for carbohydrates, lipids, and vitamin B12. Reported nutrient consumption is well below reference intakes for nearly all micronutrients, with only a small share of households achieving adequacy for niacin, vitamin A, vitamin B12, and vitamin B6. Deep shortfalls are significant across all households for calcium, folate, iron, riboflavin, thiamin, vitamin C, and zinc, with median levels below 50% of the RNI for all income quintiles. Macronutrient consumption is closer to adequacy than for many micronutrients, but median consumption of protein is below reference levels for most households, while median consumption of lipids is below reference levels for all but the two highest quintiles. Reported consumption of carbohydrate is highest for the lowest income quintiles but all households achieve adequacy for this food group. For nutrients whose median consumption is universally low across all income groups, households who cannot afford a healthy diet report median intake levels that are roughly similar to average intakes in the bottom income quintile, while in other cases adequacy is slightly higher, as for niacin, protein, thiamin, vitamin B6, and zinc. Median consumption of vitamin A, vitamin B12, and vitamin C is lower for these poorest households.

The observed range of food group adequacy is similarly wide across income quintiles, especially for fruits and animal source foods. However, shortfalls from recommendations are highest for households that cannot afford a healthy diet for all food groups except starchy staples. For all other food groups, there is a relatively large gap between food group adequacy for households that cannot afford a healthy diet and those marginally better off who are still in the lowest income quintile. Consumption of legumes, nuts, and seeds is inadequate at all income levels at a median value of less than 25% of the reference level, with little improvement as incomes rise. Consumption of fruits and vegetables is consistently below reference levels, but fruit consumption increases as incomes rise, while median vegetable consumption stays flat near 50% of the reference level. Consumption of animal source foods approaches adequacy for the upper income quintiles, and consumption of oils and fats is near complete adequacy for all income quintiles. Starchy staples consumption is highest for the lowest income quintile and declines as incomes rise, with median consumption in the highest quintile at 75% of the reference level.

**Figure 5. Reported nutrient and food group consumption relative to reference intakes by income quintile**

Panel A: Nutrient consumption

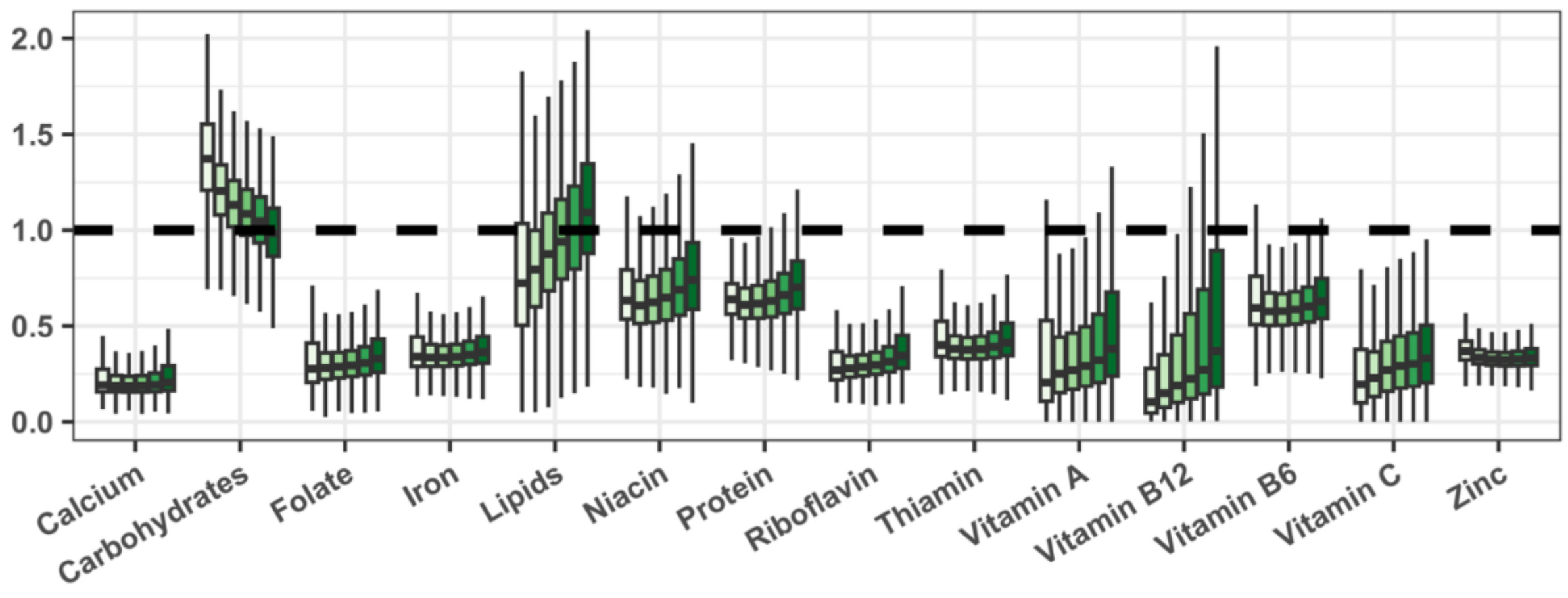

Panel B: Food group consumption

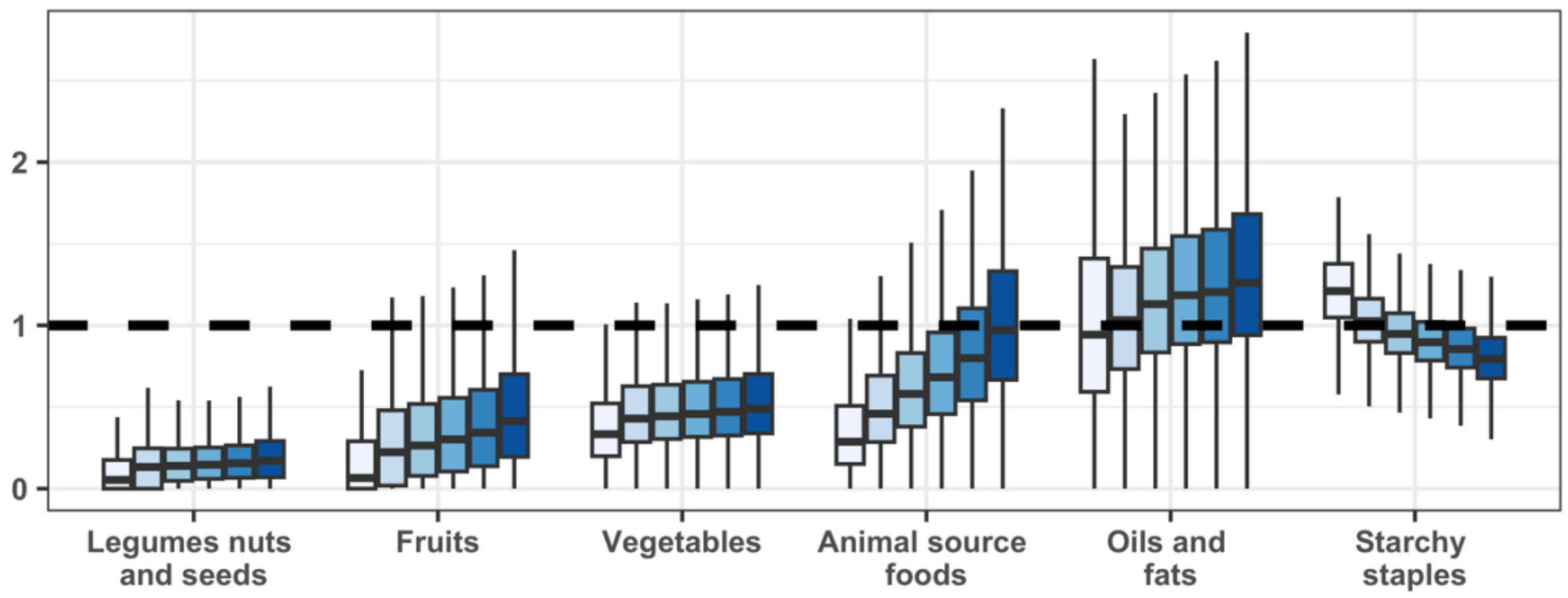

*Note: Data shown in Panel A are the ratio of reported intakes to Harmonized Nutrient Reference Values for 11 micronutrients and 3 macronutrients. Data shown in Panel B are the ratio of reported intakes to reference values from the Indonesian dietary guidelines across 6 recommended food groups, excluding*

*discretionary foods and mixed dishes. Dashed horizontal lines represent the threshold for meeting 100% of the reference value.*

The mean nutrient and food group adequacy of reported diets also varies with household income, as shown in Figure 6. Households that cannot afford a healthy diet have the worst dietary adequacy, with reported consumption meeting about 46% of both nutrient needs and food group needs. The larger subset of all households in the lowest income quintile has slightly higher dietary adequacy, meeting about 46% of nutrient requirements and 54% of food group recommendations. Dietary adequacy is not much improved even for households in the highest income quintile, who report consumption that meets 53% of nutrient requirements while providing 62% of food group needs. This modest gradient indicates that while higher incomes do help households achieve diets that are both more nutrient-adequate and closer to fulfilling dietary guidelines, these improvements are small enough that nutrient and food group adequacy of reported diets remain low.

**Figure 6. Mean nutrient and food group adequacy of reported consumption by income group**

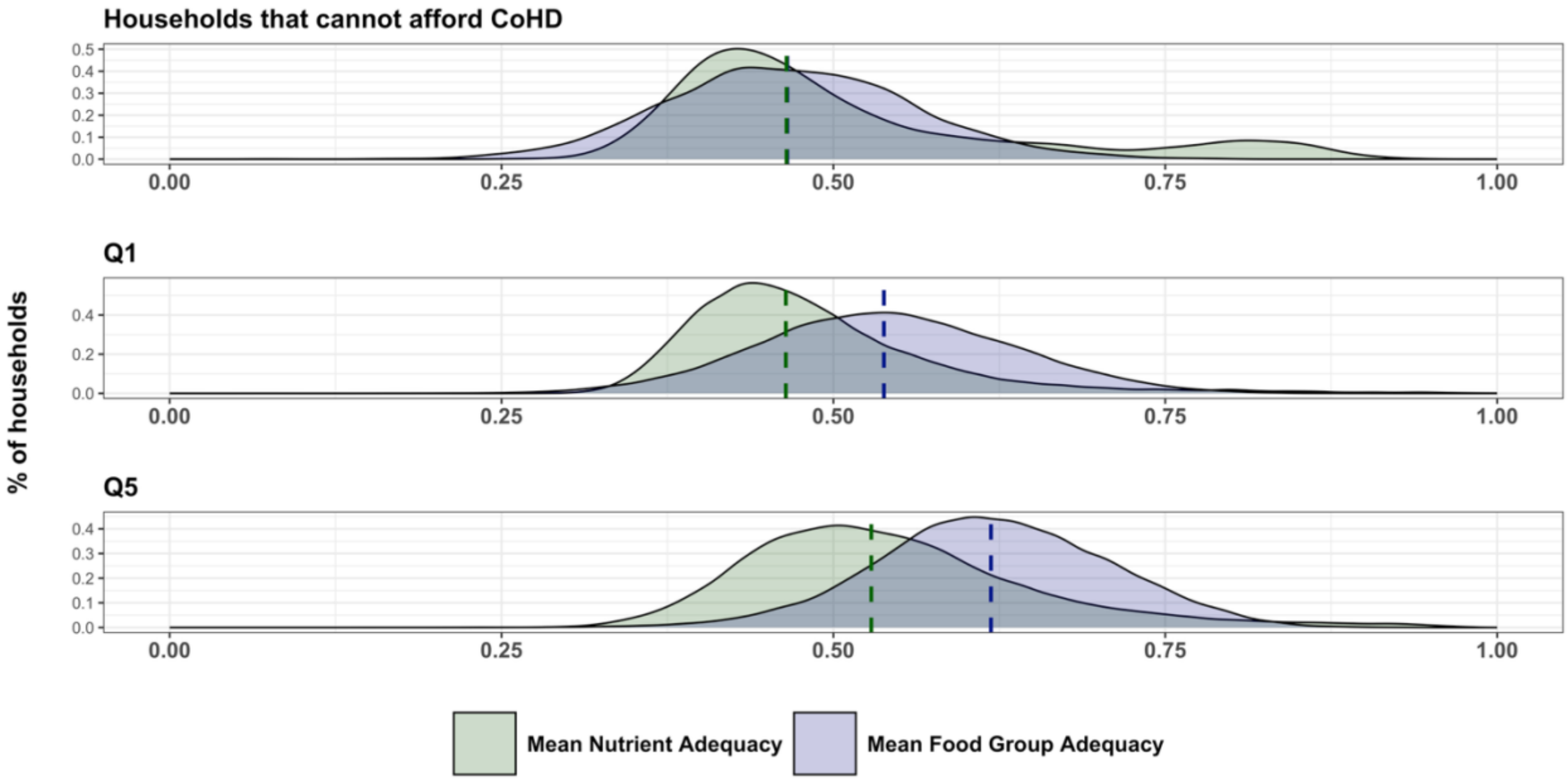

*Note: Data shown are Mean Nutrient Adequacy across 11 micronutrients and 3 macronutrients, and Mean Food Group Adequacy across 6 recommended food groups, excluding discretionary foods and mixed dishes. Dashed vertical lines represent the median score for each indicator within the expenditure quintile.*

### *3.4.Phase 3 (c): Item selection and dietary adequacy*

Income-driven shifts in nutrient adequacy and food group adequacy of reported consumption reflect substitutions within food groups as well as shifts in the relative energy share consumed from each food group. While dietary guidelines do not distinguish between recommended items in each healthy food group, there may be cost or nutrient density tradeoffs implicated in shifting item selection. Figure 7 compares the food group energy shares from least-cost healthy diets with those reported as consumed at the item-level, comparing households that cannot afford a healthy diet and those in the highest income quintile. These results are presented in terms of percentage change by item since diets are energy-adjusted to a uniform total intake of 2330 kilocalories.

The items that households report consuming are typically not the least expensive options in each food group for the modeled benchmark healthy diet, as shown in Figure 7. Within comparatively inexpensive food groups, item selection does not vary substantially between the poorest and wealthiest households. For example, reported starchy staples consumption is consistently dominated by rice regardless of income level. Similarly, consumption patterns for legumes, nuts, and seeds are highly consistent across income, with a majority of energy from tempeh and tofu.  In more expensive food groups, higher income is associated with distinct patterns of swapping energy for more preferred foods. The poorest households consume a larger share of energy from bananas and leafy greens, while the wealthiest households shift consumption away from these foods and towards other fruits and vegetables. Among animal source foods, the poorest households consume more fish and seafood, while wealthier households increase their consumption of chicken and milk products. Among mixed dishes, the poorest households consume a wide variety of mixed dishes classified as “other”, which includes fried foods, dumplings, soups, and other dishes, while the wealthiest households consume more rice-based dishes. Instant noodles account for a moderate share of discretionary food consumption, but may also be included in mixed dishes featuring noodles or in rice dishes to which noodles may be added as a source of flavor. Finally, the poorest households consume more sugar, while wealthier households consume more prepared desserts and pastries, categorized as “other sweets”.

**Figure 7. Item selection by food group in least-cost healthy diets vs. reported consumption**

**A. Least-cost healthy diets** | **B. Reported consumption (households that cannot afford a healthy diet)** | **C. Reported consumption (highest income quintile)**

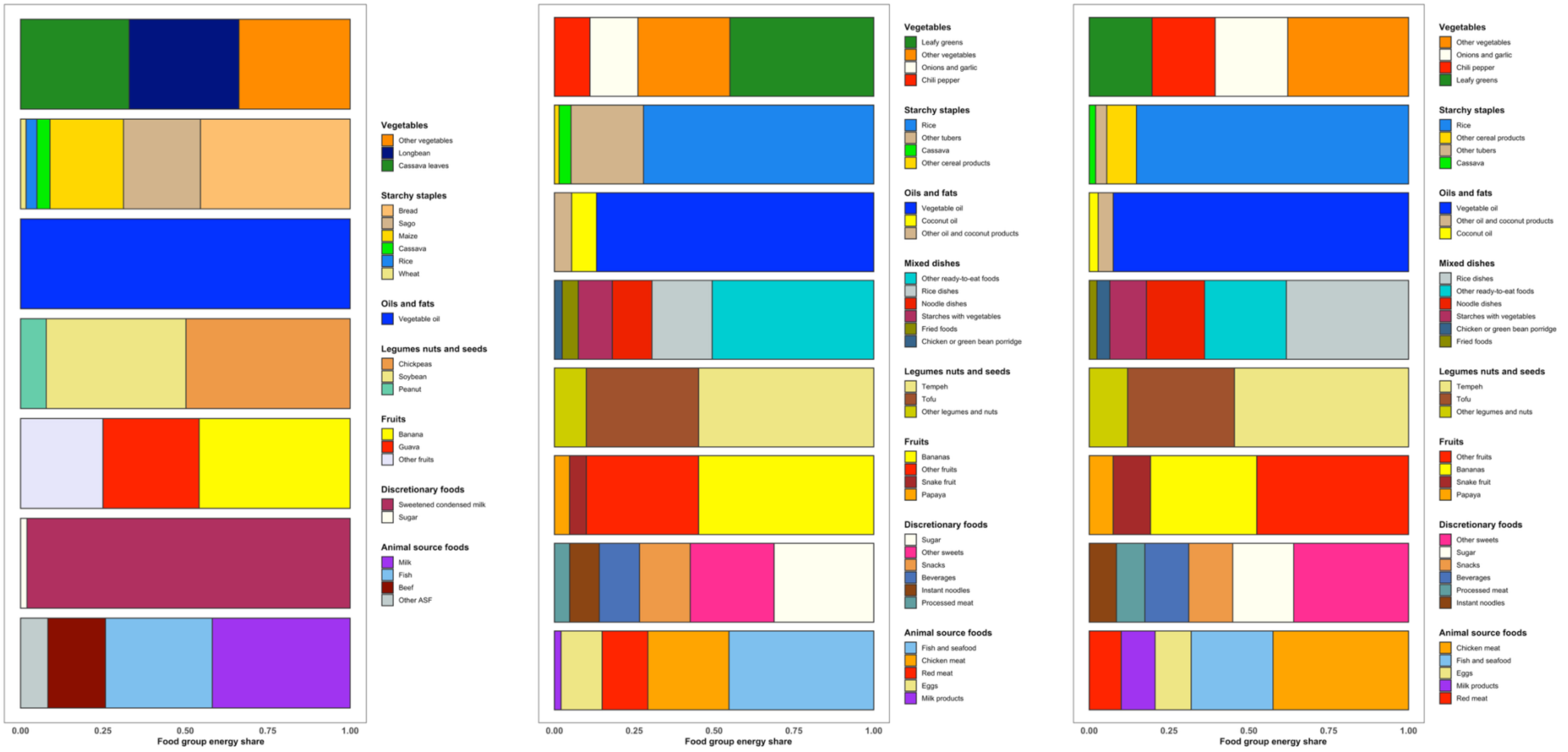

*Note: Data shown are mean energy shares for items by food group as identified in least-cost healthy diets (Panel A) or as reported for consumption (Panels B & C). Energy shares for reported consumption are calculated using household sampling weights.*

Comparing item selection in reported versus benchmark least-cost diets in Figure 7 reveals strong preferences across the income spectrum for foods that may be highly nutritious but not the least-cost options in their food group, such as tempeh and tofu in the legumes category, or chicken and eggs among animal source foods. Preferred options such as tempeh and tofu are protein-rich plant source foods that are relatively inexpensive compared to animal source foods, but are comparatively expensive relative to less processed legumes. In cases where consumption patterns do not vary with income, this may be due to strong cultural preferences, as in the case of tempeh, tofu, and rice. At the same time, households report high consumption of sweets, instant noodles, and fried foods that displace healthier options, meeting dietary energy needs that could otherwise be provided by more nutrient-rich fruits, vegetables, and legumes, nuts, and seeds. Figure 7 also illustrates how higher dietary diversity is not necessarily accompanied by high dietary adequacy. On average, modeled least-cost healthy diets include fewer foods within each food group, but much of the diversity in reported consumption comes from sweets, snacks, and packaged or ready-to-eat foods such as fried foods.

## 4. Discussion and conclusion

This study reveals where and for whom benchmark healthy diets would be affordable in Indonesia, and how reported consumption differs from the lowest-cost locally available options that would meet national dietary guidelines. Our cross-sectional findings for March 2023 reveal that Indonesia features wide geographic variation in diet costs, with costs and unaffordability highest in more remote regions, and especially Papua. High food group costs for fruits and vegetables in Papua and other high-cost regions may be best resolved through increased local production of these relatively perishable food groups, or through home cultivation that is not reflected in market prices. However, at the national level, food spending was below the minimum cost of meeting national dietary guidelines for only about 1.5% of Indonesians. This is consistent with Indonesia's status as an upper middle income country, and shows that low incomes are not the key obstacle to improved dietary adequacy. Our analysis of nutrient and food group adequacy simultaneously shows that reported diets fall short of recommendations for health, meeting only 49% of nutrient requirements and 58% of food group requirements on average. These results stem from imbalanced consumption across food groups, with imbalances varying but present at all levels of household income. In Indonesia, as in other countries where the nutrition transition is either underway or has functionally been completed, discretionary foods and mixed dishes high in added sugar, sodium, and oil have displaced healthier and more nutrient-dense fruits, vegetables, and legumes (Mehraban & Ickowitz, 2021; Nurhasan et al., 2024). While households at higher income levels have diets that come closer to meeting minimum requirements for both nutrients and healthy food groups, these households also have higher consumption of mixed dishes and discretionary foods, suggesting that while their consumption of health-promoting dietary factors is relatively higher, their consumption of health-harming dietary factors such as added sugar, sodium, and fried foods is also likely to be higher.

This study is the first to use the CoAHD indicators to illustrate how unhealthy food displaces healthier foods in countries that are undergoing the nutrition transition. The results reported here can help guide interventions to improve diets in Indonesia as well as in other countries where the nutrition transition is ongoing. Programmatic efforts in upper-middle income countries can be enhanced by understanding how and why consumers make choices that diverge from nutritional recommendations. By understanding the relationship between income and diet quality and exploring how reported consumption diverges from modeled healthy benchmarks, program designers can develop more targeted interventions that address the context-specific barriers to healthy eating. Our central finding that may be useful for program design is that for all households in Indonesia, across the income spectrum but especially for households with higher incomes, the least-cost locally available options for a healthy diet are displaced by more expensive and often less healthy items. This displacement likely occurs for reasons grounded in local and national food environments, such as culinary traditions, convenience, time use, meal preparation costs, and changes in preferences due to advertising, marketing, and labeling (Gaupholm et al., 2022). We also show that households classified as unable to afford a healthy diet as well as those in the bottom income quintile still report spending on discretionary foods and mixed dishes. This suggests that households at all income levels could bring consumption into closer alignment with nutrient and food group recommendations by reallocating existing spending toward healthy foods currently consumed in a shortfall. Policymakers developing nutrition interventions in Indonesia and other middle-income countries should rely on these insights and other findings from the three-phase analysis demonstrated here in identifying whether addressing cost, incomes, or other drivers of food group imbalance (e.g., social desirability) would lead to the greatest improvement in consumption of healthy diets in a given location.

This initial study of diet costs in Indonesia is subject to several limitations. First, our focus is purely cross-sectional, matching market prices to household consumption for a single month. Studies elsewhere address seasonal variation and trends in prices and incomes over time, which could be of great value for Indonesia as well. In addition, other estimates of CoAHD, such as those reported by the FAO and the World Bank, reflect different food group targets, geographic resolution, food price data, and estimates for household food expenditures, and thus are not directly comparable with the findings reported here. Second, our data on food composition is only approximate, especially for foods and commonly consumed mixed dishes whose nutrient density can vary widely. Even the food group composition of reported diets can be ambiguous, as the ingredients of mixed dishes and prepared meals include varying proportions of key food groups such as starchy staples and vegetables. Third and most importantly, this study offers only an initial comparison of benchmark healthy diets to reported food choice. Interventions to lower diet costs, improve affordability, and improve dietary adequacy would need to be guided by further study of the supply chains for each type of healthy food, available options for social protection and income growth, and drivers of food choice beyond income and price. Although our focus is on the majority of households whose incomes are sufficient to purchase a healthy diet, there are still low-income people in Indonesia for whom interventions to improve affordability may be needed. In places where prices for even the least expensive options are unusually high, such as Papua, investments to improve logistics, increase market coverage, and lower supply costs could bring prices towards the lower bounds observed elsewhere in Indonesia. For households with unusually low incomes, actions to improve earnings or safety net transfers would be needed to relax affordability constraints.

The findings from this study have important implications for designing targeted nutrition programs. The three-step approach demonstrated here reveals that different interventions are needed to improve diets depending on the underlying causes of poor diets. Global and national efforts to monitor the cost of healthy diets and to improve the affordability of key recommended food groups can help improve dietary adequacy and food security by helping policymakers understand these underlying causes. There are recent examples of the policy-relevance of these new tools for food security measurement. For example, national monitoring of the cost and affordability of healthy diets in Nigeria contributed to a 2024 increase in the country's minimum wage (Herforth et al., 2024). In Indonesia and other countries where affordability of healthy diets is high, the causes of poor diets are more explicitly rooted in the displacement of healthy foods. There are numerous reasons to improve household income so that all people can access the goods and services needed to live the lives they desire. However, even when the cost of nutritious foods has been brought down to the frontier of potential costs, and even when incomes are ample for meeting a locally preferred standard of living, evidence shows that people still do not consume healthy diets. This analysis has provided insights into how reported diets diverge from recommendations across the income spectrum in Indonesia, but further research will be necessary to understand how changes to the food environment, including food marketing, advertising, and labeling as described in Turner et al. (2018), might direct consumer preferences and food choice toward better alignment with benchmarks for health. Targeted interventions aimed at increasing the consumption of healthy diets have not yet been implemented in Indonesia (Huse et al., 2022). Potential policy changes that have been successful in other countries include regulatory interventions to either improve the nutrient profile of foods or to guide consumer choice away from foods and nutrients of concern. In Chile and the United Kingdom, for example, efforts to more closely regulate food marketing and product formulation have achieved notable success (Hashem et al., 2019; Rebolledo et al., 2025; Taillie et al., 2021, 2024), while taxes on sodas and other discretionary foods in Mexico led to product reformulation but had only mixed results for consumption of targeted nutrients (Aguilar et al., 2021; Salgado et al., 2025). A broader toolkit of food environment interventions including both high tax rates and consumer subsidies may be necessary to prompt product reformulation while also shifting consumer choice (Pineda et al., 2024). Other interventions aimed at incentivizing consumption of fruits, vegetables, and legumes, nuts, and seeds, will be essential to create demand for these under-consumed food groups. Least-cost diet modeling techniques may continue to provide new insights into how consumer choice diverges from recommendations in wealthier countries, and will help bring about food system transformation that enable consumers at all levels of income to choose diets that ultimately produce health.